\documentclass[acmsmall,screen]{acmart}
\setcopyright{none}
\copyrightyear{2025}
\acmYear{2025}
\acmDOI{}

\usepackage{fancyhdr}
\usepackage{graphicx} 
\usepackage{fancyvrb}
\usepackage[frozencache,cachedir=.]{minted}
\usepackage{subcaption}
\usepackage{hyperref}

\nocite{*}
\newcommand{\OurTool}{the CoCompiler}
\newcommand{\OurToolCap}{The CoCompiler}
\newcommand{\OurLanguage}{Walrus}
\newcommand{\clightgen}{\texttt{clightgen}}
\usepackage{natbib}
\title{\OurToolCap{}: DSL Lifting via Relational Compilation }
\protect\thanks{This material is based upon work supported by the Defense Advanced Research Projects Agency (DARPA) and Naval Information Warfare Center Pacific (NIWC Pacific) under Contract No. N66001-21-C-4023.  Any opinions, findings and conclusions or recommendations expressed in this material are those of the author(s) and do not necessarily reflect the views of DARPA and NIWC Pacific.}
\author{Naomi Spargo}
\affiliation{%
  \institution{Galois, Inc.}
  \city{Arlington, VA}
  \country{USA}
}
\email{nspargo@alumni.cmu.edu}

\author{Santiago Cuéllar   }
\affiliation{%
  \institution{Galois, Inc.}
  \city{Arlington, VA}
  \country{USA}
}
\email{santiago@galois.com}

\author{Jonathan Daugherty}
\affiliation{%
  \institution{Galois, Inc.}
  \city{Portland, OR}
  \country{USA}
}
\email{jtd@galois.com}

\author{Chris Phifer      }
\affiliation{%
  \institution{Galois, Inc.}
  \city{Portland, OR}
  \country{USA}
}
\email{cphifer@galois.com}

\author{David Darais      }
\affiliation{%
  \institution{Galois, Inc.}
  \city{Portland, OR}
  \country{USA}
}
\email{darais@galois.com}

\begin{document}

\settopmatter{printacmref=false}
\settopmatter{printfolios=true}
\renewcommand\footnotetextcopyrightpermission[1]{}
\pagestyle{fancy}
\fancyfoot{}
\fancyfoot[R]{miniKanren'25}
\fancypagestyle{firstfancy}{
\fancyhead{}
\fancyhead[R]{miniKanren'25}
\fancyfoot{}
}
\makeatletter
\let\@authorsaddresses\@empty
\makeatother
\begin{abstract}
Lifting low-level or legacy code into a domain-specific language (DSL) improves our ability to understand it, enables deeper formal reasoning, and facilitates safe modification. We present \OurTool{}, a bidirectional compiler and lifter between C and Lustre, a synchronous dataflow language used for reactive systems \cite{lustre}. The key insight behind \OurTool{} is that writing a compiler as a relation, rather than as a traditional function, yields a DSL lifter “for free”. We implement this idea by re-encoding the verified Lustre-to-C compiler Vélus in the \OurLanguage{} relational programming language. This solves what we call the vertical lifting problem, translating \emph{canonical} C into Lustre. To address the complementary horizontal problem—handling real-world C outside the compiler's image—we apply semantic-preserving canonicalization passes in Haskell. The resulting tool, \OurTool{}, supports lifting real reactive C code into Lustre and onward into graphical behavioral models. Our approach is modular, language-agnostic, and fast to implement, demonstrating that relational programming offers a practical foundation for building DSL lifters by repurposing existing compilers.
\end{abstract}

\maketitle
\thispagestyle{firstfancy}

\section{Introduction} 
\label{sec:introduction}

\subsection{The DSL lifting problem}
\label{sec:DSLs}
Domain Specific Languages are powerful abstractions that make domain concepts first-class, allowing developers to write clearer, safer, and more analyzable code within their specific problem space. For example, the Lustre language is tailored for \emph{reactive systems}, providing specialized abstractions for time and concurrency, along with a rich ecosystem of analysis tools like Kind2 that help developers verify correctness \cite{lustre, kind2}. However, many reactive systems are still written in low-level languages like C, where high level domain structure is obscured by implementation details. Lifting these systems into Lustre exposes their underlying intent and structure, making them easier to understand, adapt, maintain, or prove correct~\cite{velusBlog}.

Unfortunately, DSL lifting is not an easy task. Although Lustre~$\rightarrow$~C compilers abound, C~$\rightarrow$~Lustre decompilers are rare and challenging to build. This asymmetry is the inspiration for our work: we demonstrate an unusual approach to quick and easy DSL lifting and a surprising application of relational programming with \OurTool{}: a bidirectional Lustre~$\leftrightarrow$~C compiler and lifter. CoCompiler users can lift their reactive systems into Lustre, work on them in a conducive environment, and then compile them back down into C. Our secondary motivation for creating \OurTool{} was to demonstrate that relational programming simplifies many challenges inherent to DSL lifting and language translation. 

The problem of lifting low-level code into a DSL has two orthogonal components. The first is \textit{\textbf{vertical translation}}: identifying and recovering high-level DSL abstractions from C code that clearly expresses domain-relevant behavior, albeit in low-level terms. This involves moving up in abstraction from a well-understood \emph{canonical} sublanguage of C into the DSL. The second is \textit{\textbf{horizontal transformation}}: many C programs that morally belong to the domain (e.g. represent reactive systems) are not written in a way that makes their structure immediately liftable. The goal of the horizontal transformation stage is to recognize idiomatic C that has domain-specific structure, then rewrite this C into the canonical form that makes this structure explicit. Together, these two transformations form a full DSL decompiler. 

\subsection{\OurToolCap{}}
\label{sec:OurTool}
\OurToolCap{} is a C~$\rightarrow$~Lustre 
lifter that similarly partitions the lifting task into a \textit{\textbf{vertical translation}} and \textit{\textbf{horizontal transformations}}. We present \OurTool{} as a novel approach for efficiently solving the vertical translation problem with minimal expertise in either the source or target language. While \OurTool{} includes a few example-driven horizontal transformations, these are currently very limited. As of this paper, \OurTool{}'s ability to recognize reactive systems is insufficient to lift idiomatic C programs. Broadening the range of reactive C programs \OurTool{} can recognize, as well as designing a general technique for horizontal transformation, is an open direction for future work.

Concretely, the vertical part of \OurTool{} is a relational implementation~\cite{miniKanren, reasonedSchemer} of an existing functional Lustre to C compiler. The relational approach to vertical translation has multiple advantages. First, it reduces the vertical translation problem to Lustre~$\rightarrow$~C compilation. To see this, consider a \OurLanguage{} relation called \texttt{compile} that relates Lustre and C programs. \OurLanguage{} is embedded in Haskell and in this embedding \texttt{compile} has type (roughly): 
\begin{verbatim}
compile :: Lustre -> C -> Goal ()    
\end{verbatim}
The \texttt{Goal ()} return type represents the result of \OurLanguage{} solving constraints on the two arguments \cite{walrus}.
We can use \texttt{compile} by passing it a Lustre program and a C program; the \OurLanguage{} relational programming engine will check if they are related by compilation (i.e. if the Lustre compiles to the C). More interestingly, we can pass a concrete program for one argument and an unbound logical variable for the other. Then, \OurLanguage{} will solve for the variable as it solves \texttt{compile}'s constraints. Crucially, if we pass a variable for the Lustre argument and concrete C program for the C argument,  \texttt{compile} will solve for Lustre that satisfies the compilation relation.

Using miniKanren syntax, this is how to use  \texttt{compile} as a compiler and as a lifter \cite{reasonedSchemer}. If variable \texttt{cFile} is fresh and variable \texttt{lustreFile} is already bound to a Lustre AST, running:
\[ \texttt{(run 1 (cFile) (compile lustreFile cFile))} \] will produce the compiled C AST for \texttt{lustreFile}. Now, say \texttt{lustreFile} is a fresh variable and \texttt{cFile} is already bound to a C AST. Say also that \texttt{cFile} is in the image of \texttt{compile}. Running:
\[ \texttt{(run 1 (lustreFile) (compile lustreFile cFile))} \] will produce a Lustre AST that, when compiled, results in \texttt{cFile}. This achieves our desired \textit{vertical translation} (as defined in section \ref{sec:DSLs}). Using this approach, we can lift everything in \texttt{compile}’s image, and we identify this image as the “well-understood, \textit{canonical} sublanguage” that our vertical translation supports.

We based our implementation of \texttt{compile} on a mature, verified, functional compiler called Vélus, which provided us with a solid foundation and blueprint for $\OurTool{}$'s design ~\cite{velusRepo, velusOG, velusN}. By porting  Vélus into the relational setting, we obtain a relational lifter with little effort, since any C program in Vélus's image can be lifted back to its corresponding Lustre representation. 

This implementation strategy reveals the second advantage of our approach: it is language and compiler-agnostic. We were able to identify canonical C and lift it into Lustre without having to understand the synchronous reactive semantics of the C we lifted, or even of the Lustre language. We didn’t even need to fully understand the transformations performed by the compiler we ported. 

Thanks to our approach’s simplicity and its reuse of existing, trustworthy code, \OurTool{} was relatively quick to implement. We spent just over three engineer-weeks porting the core compilation logic from Vélus.\footnote{Specifically, we ported four compiler passes, each taking approximately 32 hours, including time spent advancing \OurLanguage{}.} However, we developed \OurLanguage{} in parallel with \OurTool{}. Because of this, \OurTool{} developers wrote substantial boilerplate by hand, debugged without \OurLanguage{} debugging support and often stopped to improve \OurLanguage{}'s infrastructure and standard library. Now that \OurLanguage{}'s foundations are in place, we estimate that similar \OurTool{} development efforts could be completed in half the time.

The main disadvantage of this approach is that the range of programs we can lift starts out very limited. Our vertical translation can \textit{only} lift C programs \textit{exactly} in Vélus's image. Any syntactic change that moves a program outside Vélus's image will make that program unliftable by \texttt{compile}. We have slightly broadened the range of liftable programs with some \textit{horizontal transformations}. 


We think that more X$\leftrightarrow$~Y bidirectional compiler/lifters can be built with our approach: first port an existing non-relational X$\rightarrow$~ Y compiler to the relational setting, then connect Y$\rightarrow$Y horizontal transformations to broaden the range of liftable programs. We present \OurTool{} as the first application of our approach to Lustre~$\leftrightarrow$~C compilation/lifting and to bidirectional compilation generally. We have not yet applied this approach to any other lifting/compilation problems.

Concretely, the contributions of this short paper are as follows:
\begin{itemize}
\item We present a novel technique for rapidly prototyping DSL lifters by using relational languages to build bidirectional compilers.
\item We demonstrate the technique with \OurTool{}, a C~$\leftrightarrow$~Lustre compiler and lifter. \OurToolCap{} is comprised of (1) a relational, bidirectional version of the Vélus functional compiler, (2) simple, semantics-preserving, single-direction transformations from a broader subset of C into Vélus’s image, and (3) semantics-preserving, single-direction translations from Lustre into graphical SCADE. 
\item We demonstrate the usability of \OurTool{} as a full DSL lifter from C to SCADE block diagram.
\end{itemize}

The rest of the paper is structured as follows. In Section~\ref{sec:related work}, we describe work that inspired our use of relational programming or solved similar DSL lifting problems with different techniques. Then, Section~\ref{sec:background} gives an overview of the languages and tools we used when building \OurTool{}. In Section~\ref{sec:ourTool}, we give a thorough presentation of the concept and implementation of \OurTool{}, as well as an example of what \OurTool{} produces from a real C file. Finally, we summarize our work and discuss future research directions in Section~\ref{sec:conclusion}.


\section{Related work}
\label{sec:related work}
Our approach to rapid DSL lifter prototyping builds on a long history of work in relational programming. Byrd et al.~\cite{miniKanren} demonstrated how functional programs such as interpreters can be ported to a relational setting to unlock new behaviors. Indeed, a relational interpreter can be run \emph{forward} to compute the output of a program and \emph{backward} to synthesize a program from example outputs. 
To further demonstrate the technique of repurposing functional programs in the relational setting, Byrd, Rosenblatt, and others~\cite{barliman, barlimanRepo} designed Barliman: a prototype “smart editor” that synthesizes programs based on user-provided tests. The crucial insight behind this work is that a program synthesizer can be thought of as the inverse of a program interpreter, so to create a program synthesizer, one need only write a relational interpreter and run it backwards. This insight inspired us to create a DSL lifter by writing a DSL compiler. In both circumstances, we find a pair of problems (interpreter and synthesizer, compiler and lifter) where one element of the pair is harder than the other. The relational setting allows us to exploit this asymmetry. 

 There have been many other relational decompilers in recent years: GrammaTech’s Datalog-based disassembly framework \cite{datalog}, Gigahorse (a decompiler for Ethereum smart contracts based on Soufflé) \cite{gigahorse}, and Securify2 \cite{securify2} all express the semantics of compiled code as logical constraints. These tools leverage Datalog’s ability to naturally express constraints typical of decompilation. But, while they have relational semantics, they are directly designed as decompilers and can’t be executed “forwards”. To the best of our knowledge, \OurTool{} is the first system to derive a relational decompiler from a functional compiler, and to support compilation and decompilation. We gained the decompilation ability by working in a relational setting, but we also retained the ability to compile. 

\OurToolCap{} is one of many diverse C~$\rightarrow$~Lustre lifters; all these tools try to solve the same problem, but use widely varying approaches. Blanc et. al. present  Frama-C/Synchrone, a C~$\rightarrow$~Lustre lifter built on the Frama-C verification framework \cite{framaC}. Like \OurTool{}, Frama-C/Synchrone recovers high-level Lustre models from C code. However, Frama-C/Synchrone was designed specifically for C~$\rightarrow$~Lustre lifting; Blanc et. al developed a sophisticated theory of Lustre semantics in terms of C constructs, then implemented this theory to solve the vertical and the horizontal problem in the C~$\rightarrow$~Lustre case. Their work is very effective, but limited to C~$\rightarrow$~Lustre lifting. \OurToolCap{} demonstrates a more generic approach to the same problem.

In a recent paper \cite{lustreToScade}, Grimm et. al. describe a further level of lifting, from text-based Lustre models to graphical representations similar to Safety Critical Application Development Environment (SCADE) models. As with Frama-C/Synchrone, this work is a bespoke solution to a particular lifting problem, rather than a technique for DSL lifting in general. \OurToolCap{} also presents a single-direction translation from Lustre to SCADE, though it is less comprehensive than that of Grimm et. al. We included our Lustre~$\rightarrow$~SCADE translation merely for completeness; we are interested in Grimm et. al’s approach and hope to improve our ability to generate block diagrams in future work.

\section{Background: languages and tools}
\label{sec:background}
\subsection{Lustre}
\label{subsec:lustre}
Lustre is a language designed for reactive systems, which are programs that receive a stream of input directly from the environment and react in real time. To provide reactive systems programmers with useful abstractions, Lustre has some unusual features. It is a synchronous dataflow language, meaning that every variable is a stream indexed by a native notion of time. Each variable has a type and a clock, which is a temporal annotation of the time-steps at which the variable’s value is well-defined. Lustre is declarative, meaning that there is no notion of an iteratively modified state. This makes Lustre easy to reason about and a great candidate for programming critical reactive systems like medical devices, sensors, and satellites. Today, there are many industrial tools for designing reliable systems, including Ansys SCADE, Kind2, JKind, and NKind target Lustre \cite{lustre, kind2, nkind}.

In Figure~\ref{fig:count.lus}, we show a simple example of a Lustre program, \texttt{count}, that takes an integer input stream \texttt{i} and outputs a stream whose value at timestep $t$ is the sum of the input's value at $t$ and all previous input values. Table~\ref{table:count} shows the input/output execution of the same example.
\
\begin{figure}
\caption{A Lustre program that adds all its inputs.}
\label{fig:count}
\begin{subfigure}[T]{.50\textwidth}
\centering
\begin{verbatim}
node count (i:int) returns (o:int)
let
  o = (0 fby o) + i;
tel

\end{verbatim}
\caption{Textual representation of \texttt{count}.}
\label{fig:count.lus}
\end{subfigure}
\begin{subfigure}[T]{.40\textwidth}
\centering
        \begin{tabular}{ccc}
            \hline Timestep & Input \texttt{i} & \texttt{o} \\ [0.5ex] 
            \hline\hline
            0 & 5 & 5\\
            1 & 4 & 9\\
            2 & 0 & 9\\
        \end{tabular}
\caption{Output stream over time for \texttt{count}.}
\label{table:count}
\end{subfigure}

\end{figure}

Lustre programs are composed of \texttt{node}s which represent units of computation. \texttt{fby} is one of Lustre's many primitives for programming reactive systems. Pronounced ``followed by'', \texttt{fby} is an infix primitive that takes two arguments and evaluates to a stream. The left argument is a constant indicating the first value of the resulting stream. The right argument is a stream indicating all subsequent values of the resulting stream. So, the expression \texttt{(0 fby o)} is a stream with value zero at timestep $0$, and the value of stream \texttt{o} at $t$ for timestep $t > 0$ . 

\begin{figure}
\centering
\includegraphics[scale=0.3]{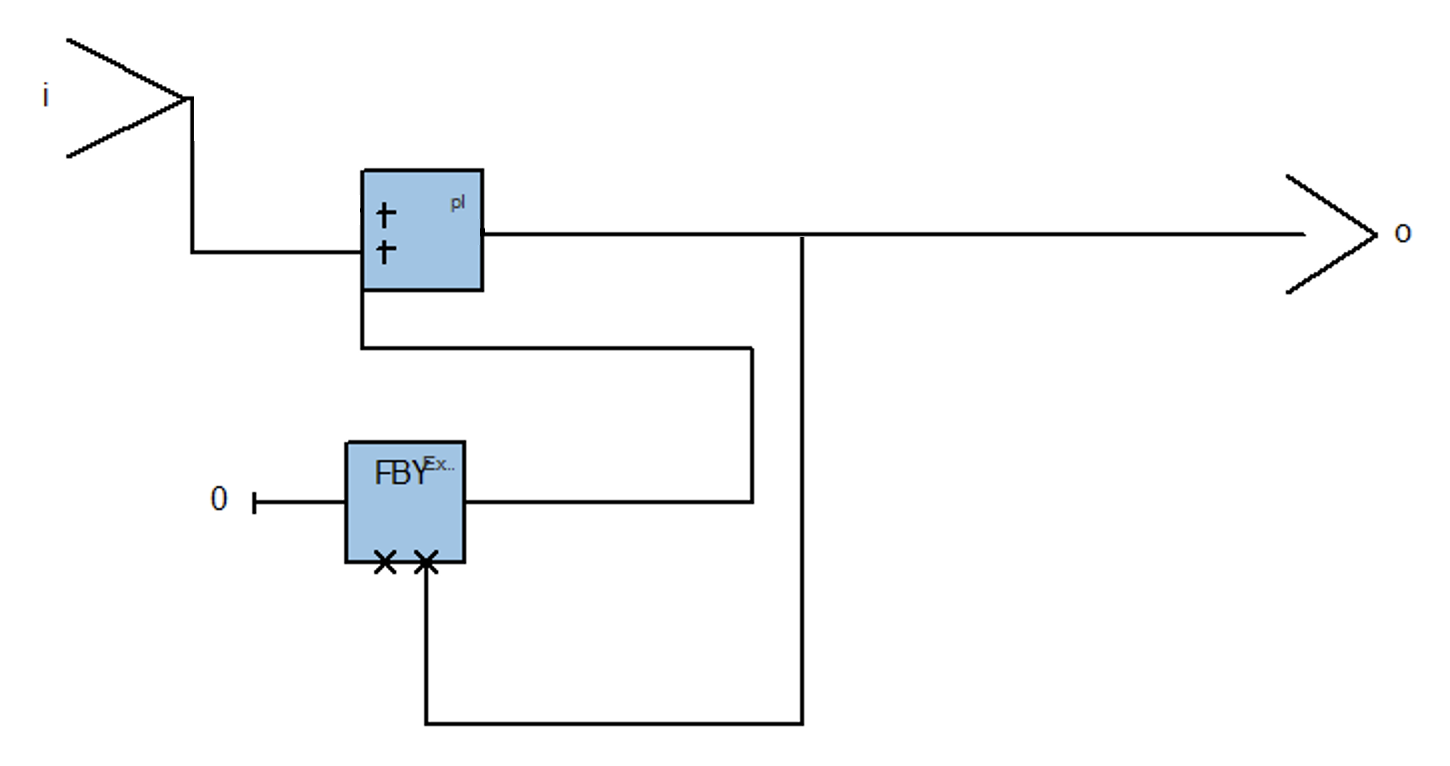}
\caption{SCADE model for \texttt{count}}
\label{fig:count-SCADE}
\end{figure}

Lustre programs can be graphically represented as block diagrams using a tool like Ansys SCADE. Indeed, Lustre was designed to provide a textual representation and executable version of the block diagrams commonly used by control systems engineers, and is sometimes referred to as a ``block diagram language'' \cite{lustreOld, velusBlog}. Figure \ref{fig:count-SCADE} is the SCADE block diagram corresponding to \texttt{count.lus}. 

There are many Lustre versions, some of which are incompatible with each other. Both Vélus and \OurTool{} operate on a sublanguage of Lustre V4 \cite{velusRepo, velusOG, velusN}. 

\subsection*{Vélus and CompCert}
\label{subsec:velus}
Vélus is a verified compiler from Lustre to Clight\footnote{Clight is a dialect of C designed for analysis and verification \cite{clight, compcertRepo}.} written in Rocq and OCaml. Vélus comes with a Rocq definition of Lustre semantics and a Rocq proof that compiled code correctly implements the original Lustre program. Vélus connects to Compcert, a Rocq-verified compiler from C to assembly, allowing the user to produce verified executables from their Lustre programs \cite{compCert, compcertRepo}. The Vélus authors describe their compiler as "as an extension of CompCert for compiling Lustre" \cite{velusWeb}.

Out of the many Lustre~$\rightarrow$~C compilers available, we chose to base \OurTool{} on Vélus because 1) Vélus’s proofs make us confident that \OurTool{} preserves semantics as well, 2) Vélus is written in a functional style, making it easier to port to the relational setting, and 3) Vélus outputs Clight, which has a thriving tool ecosystem around it that facilitated \OurTool{}'s development \cite{clight, compcertRepo, clight2}.
In particular, we used \clightgen{}, a CompCert tool that translates C into Clight, as the first step in our C~$\rightarrow$~Lustre lifting toolchain \cite{clightGenRepo}. When run on C code, \OurTool{} first calls out to \clightgen{}, then passes the resulting Clight to our implementation in Haskell and \OurLanguage{}. 

The vertical translation component of \OurTool{} is simply a partial re-implementation of Vélus in \OurLanguage{}. We changed each Rocq function into an \OurLanguage{} relation, but preserved much of Vélus's high-level design. \OurToolCap{} uses the same compilation phases, AST designs, even the same helper functions and call graphs. At the compiler design level, the only difference between \OurTool{}'s vertical component and Vélus is that we support fewer Lustre programs and omit of some of Vélus's compiler optimizations. This is only because they were not necessary for \OurTool{} to serve as a proof of concept for relational compilation and lifting.


\begin{figure}
\centering
\includegraphics[scale=0.5]{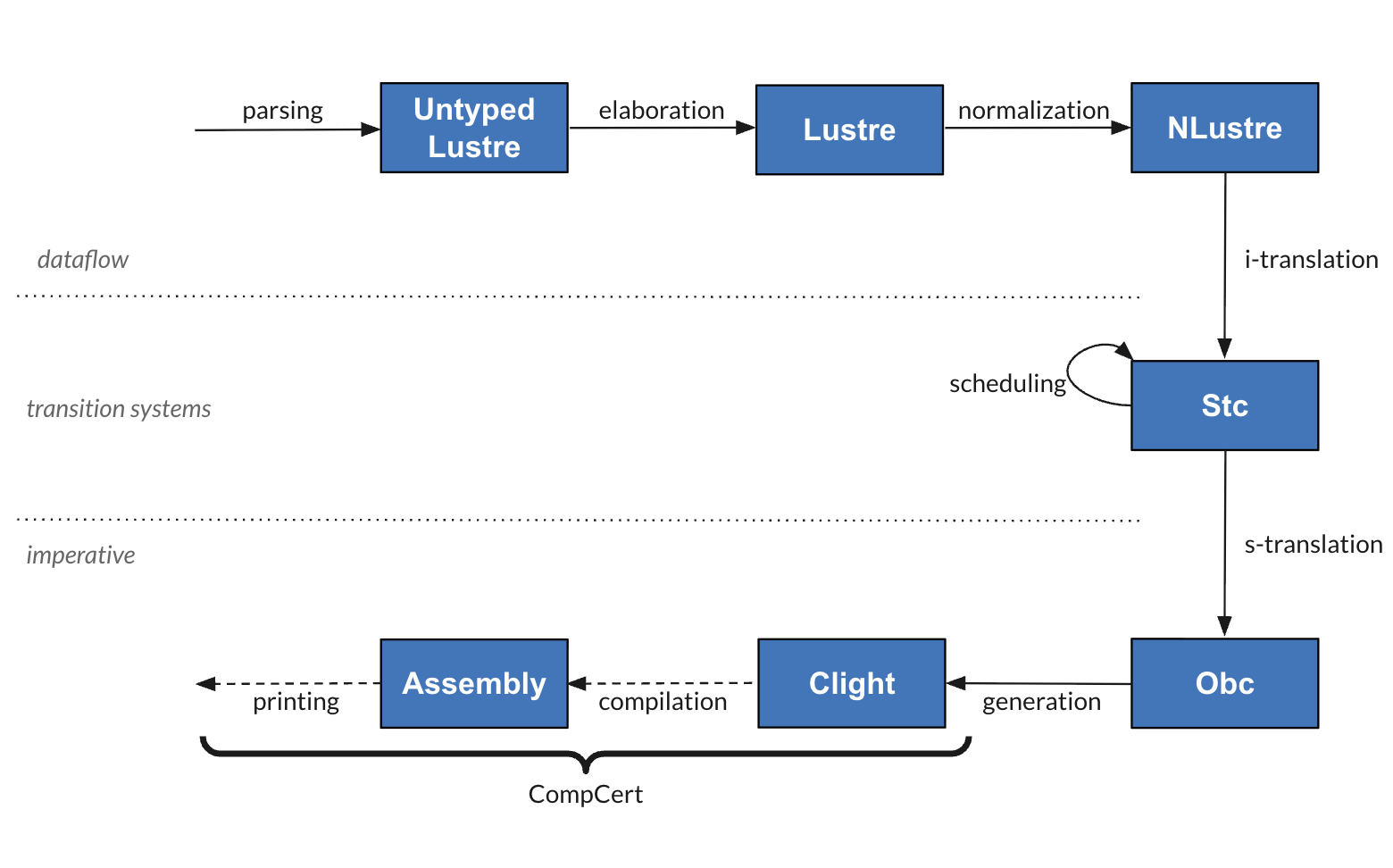}
\caption{Vélus compilation phases \cite{velusN, velusWeb}}
\label{fig:Velus}
\end{figure}

Figure \ref{fig:Velus} shows Vélus's key compilation phases \cite{velusN, velusWeb}. Lustre files are parsed, then annotated with types and clocks by the \textit{elaboration} pass \cite{velusWeb}. Then, Vélus’s first compilation phase ‘normalizes’ each Lustre program into a simplified sublanguage — \textit{normalized Lustre} (also called \textit{NLustre}) \cite{velusN}. Although \OurTool{} translates from a Lustre AST to a simplified normalized Lustre AST, it do not yet implement bidirectional normalization. All lifted Lustre programs will be in normal form, and \OurTool{} cannot compile non-normalized Lustre programs. From normalized Lustre, Vélus translates streams into state \textit{i}nstances with \textit{i-translation} \cite{velusN}. In the \textit{\textbf{S}ynchronous \textbf{T}ransition \textbf{C}ode} intermediate representation, a reactive system is represented as a composition of state transitions \cite{velusWeb}. Vélus \textit{schedules}, or reorders, the state transitions before passing to the next phase, where the native notion of time is lost and instruction order becomes significant \cite{velusWeb, velusN}. The CoCompiler omits the scheduling pass and so can only compile correctly scheduled Lustre code. This means the user must order their Lustre stream operations in a temporally sensible way. In the next phase, Vélus performs \textit{s-translation} to create an ordered \textit{s}equence of instructions that manipulate an encapsulated state \cite{velusWeb, velusN}. The result is \textit{\textbf{Ob}ject \textbf{c}ode}, imperative code with distinct \textit{step} and \textit{reset} functions that persist into the compiled Clight, as shown in Figure \ref{fig:count.c} \cite{velusWeb}. 

Each Vélus intermediate representation has syntax and semantics specified in the Rocq theorem prover, and each compilation phase is accompanied by a proof that it preserves semantics \cite{velusWeb}. 





\subsection*{\OurLanguage{} relational language}
\label{subsec:ourLanguage}
\OurLanguage{} is a miniKanren-style logic programming language shallowly embedded in Haskell. As in miniKanren, relations in \OurLanguage{} can be run to solve for any argument \cite{miniKanren, reasonedSchemer}. This capability, characteristic of relational programming, undergirds our method of creating quick and easy DSL lifters. Only in the relational setting does the decompilation problem naturally reduce to compilation.

\begin{figure}
\centering
\begin{minted}{haskell}
-- | Multiplication relation on @Nat@s.
mulR :: Nat -> Nat -> Nat -> Goal ()
mulR x y mulxy = do
  disj [ -- This disjunction corresponds to casing on @x@
     -- case where @x@ is 0
     do x === O
        mulxy === O
     -- case where @x@ is 1
   , do (x', mulxy') <- fresh2
        x === S x'
        mulR x' y mulxy'
        addR y mulxy' mulxy
    ]

-- | Principal square root relation on @Nat@s.
squareR :: Nat -> Nat -> Goal ()
squareR rt sq = mulR rt rt sq
\end{minted}
\caption{\texttt{mulR} and \texttt{squareR} relations in \OurLanguage{}}
\label{fig:squareR}
\end{figure}

To see how to use \OurLanguage{} relations, consider the simple \texttt{mulR} and \texttt{squareR} relations on unary natural numbers (Figure \ref{fig:squareR}). These relations are contained in \OurLanguage{} standard library file \texttt{Unary.hs}; though they are \OurLanguage{} relations, they are also monadic Haskell functions  \cite{walrus}. We can use \texttt{squareR} to compute squares. Running the command: \[ \texttt{(run* (result) (squareR 5 result))} \] solves for the second argument of \texttt{squareR} and results in \texttt{25}. More interestingly, we can also use \texttt{squareR} to compute square roots. Running: \[ \texttt{(run* (result) (squareR result 25))} \] solves for the \textit{first} argument of \texttt{squareR} and results in \texttt{5}. The same \texttt{squareR} code can be run forwards to compute squares or backwards to compute square roots \cite{miniKanren}. In this way, our \texttt{compile} function discussed in the introduction can be run forwards  as a Lustre~$\rightarrow$~C compiler, or backwards  as a C~$\rightarrow$~Lustre lifter.

A Haskell type becomes usable in \OurLanguage{} when it implements the \texttt{Unifiable} typeclass, thereby explaining to the \OurLanguage{} solving engine how terms of that type should be unified~\cite{walrus}. In order to implement \OurTool{}, we ported the ASTs\footnote{Lustre, normalized Lustre, Stc, Obc, and Clight} from four of Vélus’s compilation phases into Haskell and made them \texttt{Unifiable}, thereby porting them into \OurLanguage{} \cite{velusRepo, velusWeb}. We then rewrote Vélus’s compilation phases as relational programs in \OurLanguage{} to create the vertical translation part of \OurTool{}.

\section{\OurToolCap{}}
\label{sec:ourTool}

\subsection{Horizontal and vertical components}
\label{horizontalVertical}
An ideal bidirectional compiling/lifting relation captures the full semantic correspondence between source and target programs. Specifically, it relates C and Lustre programs that exhibit the same observable behavior. This relation is inherently many-to-many: a single Lustre program may correspond to many semantically equivalent C implementations, and vice versa. Superficial variations in either language, such as alpha-renaming, the use of \texttt{for} versus \texttt{while} loops, or the order of function declarations, should ideally not affect whether a C program is considered a valid implementation of a Lustre program. This ideal relation answers two largely orthogonal concerns: semantic equivalence within a language, which we refer to in Section \ref{sec:introduction} as the \emph{horizontal} problem, and compiling/lifting, which we call the \emph{vertical} problem.

\begin{figure}
\centering
\includegraphics[scale=0.4]{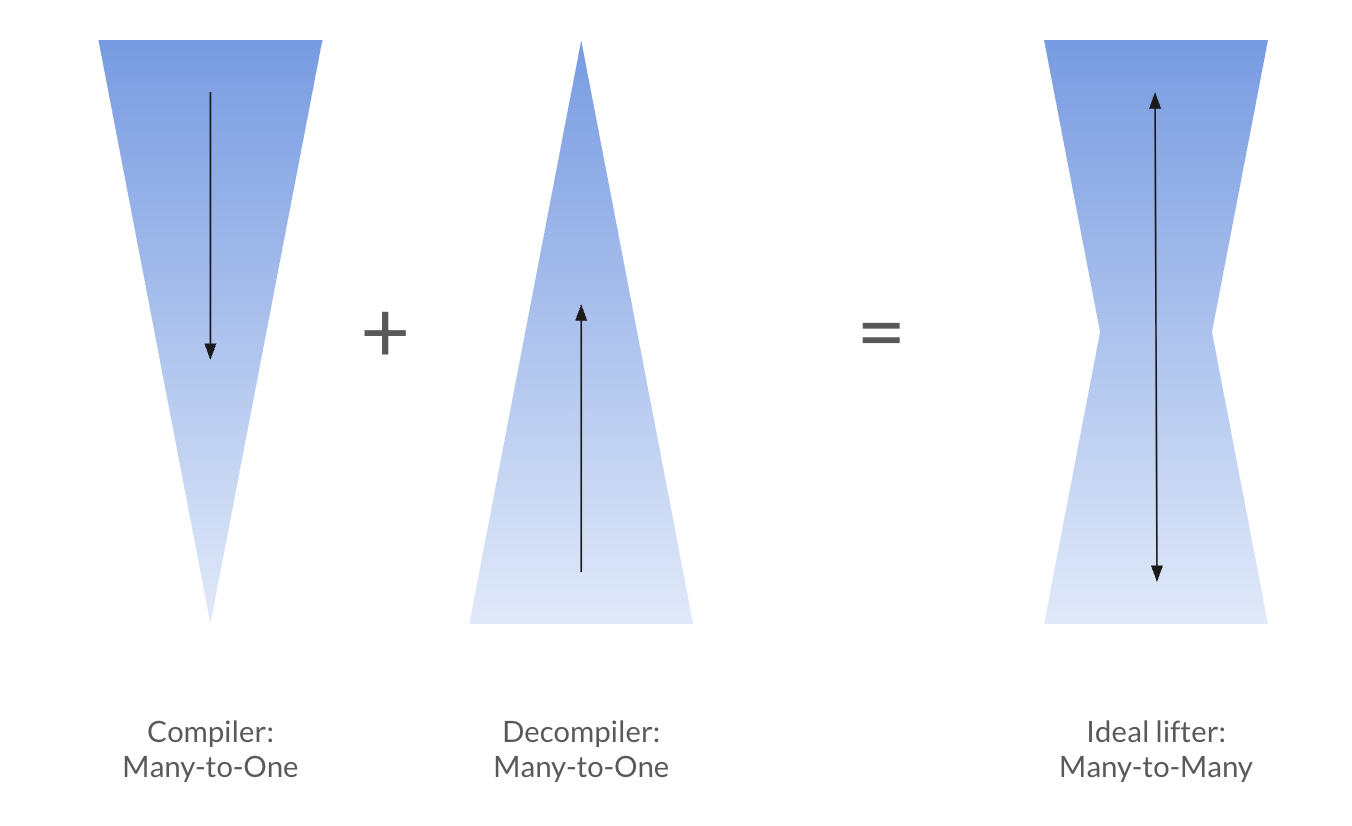}
\caption{Compiler, decompiler, and ideal lifter}
\label{fig:ideal lifter}
\end{figure}

 By contrast, all non-relational Lustre~$\rightarrow$~C compilers choose one particular C representation for each Lustre program, making them many-to-one. Indeed, Vélus has canonicalization passes that ensure many semantically equivalent Lustre programs are mapped to the same result. Similarly, a non-relational C~$\rightarrow$~Lustre decompiler, built from scratch, would also be many-to-one, but in the opposite direction. By combining the relational information contained in a compiler and in a decompiler, one could accurately capture the many-to-many relation between Lustre and C, as illustrated in Figure \ref{fig:ideal lifter}. 
 
 
Although Vélus is many-to-one and single-direction, the Vélus compilation process also separates concerns into vertical and horizontal. Vélus first normalizes a Lustre program to address the issue of semantic equivalence ~\cite{velusN, velusRepo}. Then, Vélus applies a vertical \textit{core} of mostly one-to-one transformations, resulting in compiled Clight ~\cite{velusRepo}.

\begin{figure}
\centering
\includegraphics[scale=0.4]{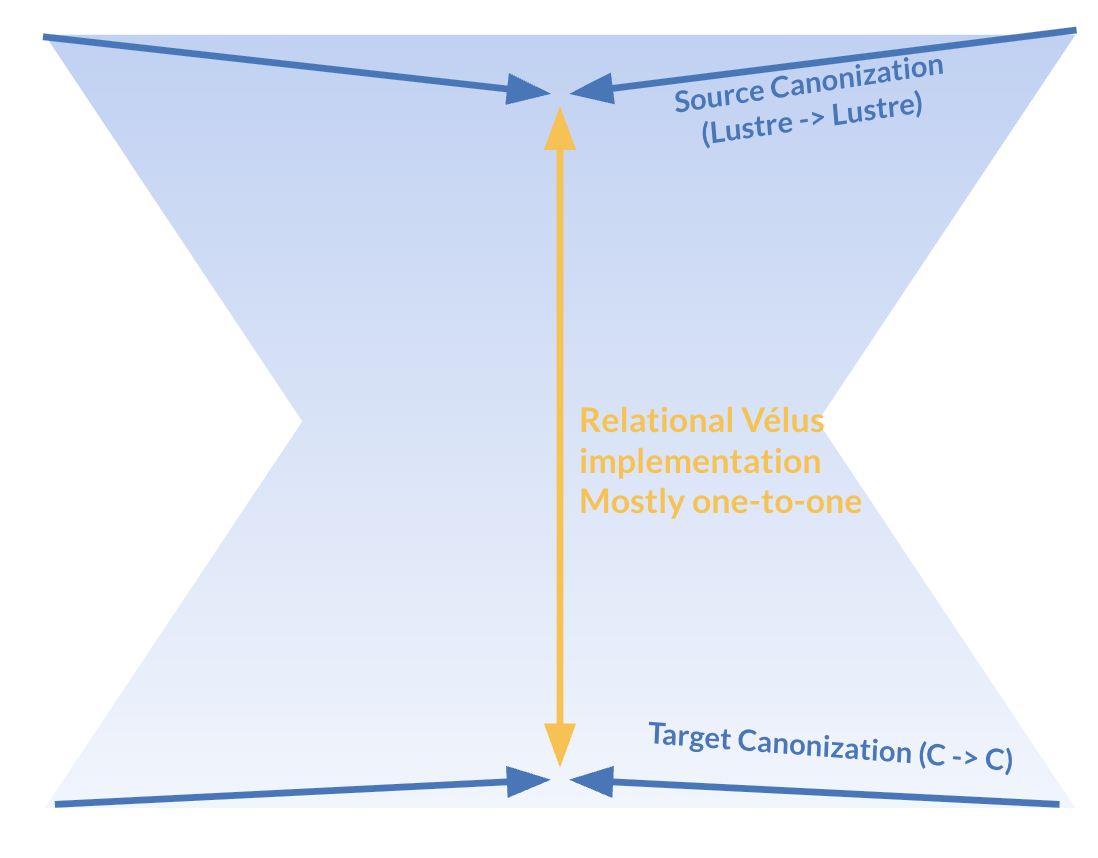}
\caption{\OurToolCap{} is ‘one to one’ + canonicalization}
\label{fig:cocompiler}
\end{figure}

 When porting Vélus to \OurLanguage{}, we focused on Vélus's core one-to-one component. Once implemented relationally, this component allows \OurTool{} to both compile and lift programs, though only within Vélus’s domain and image. We refer to these sets as the \textit{canonical} sublanguages of Lustre and C. To broaden the applicability of \OurTool{} and better approximate an ideal bidirectional compiler/lifter, we introduced a modest C-to-C canonicalization pass. While a true many-to-many relation between C and Lustre would require this pass to be bidirectional, our current implementation performs it in unidirectional Haskell code. We designed \OurTool{} to reflect the vertical/horizontal problem partition; the relational implementation of Vélus's core is the vertical translation and the canonicalization transformation runs horizontally. Figure~\ref{fig:cocompiler} illustrates how combining a relational one-to-one core with pre- and post-processing canonicalizations yields a practical approximation of an ideal bidirectional decompiler.

\subsection{Architecture}
\label{subsec:implementation}

\begin{figure}
\centering
\includegraphics[scale=0.42]{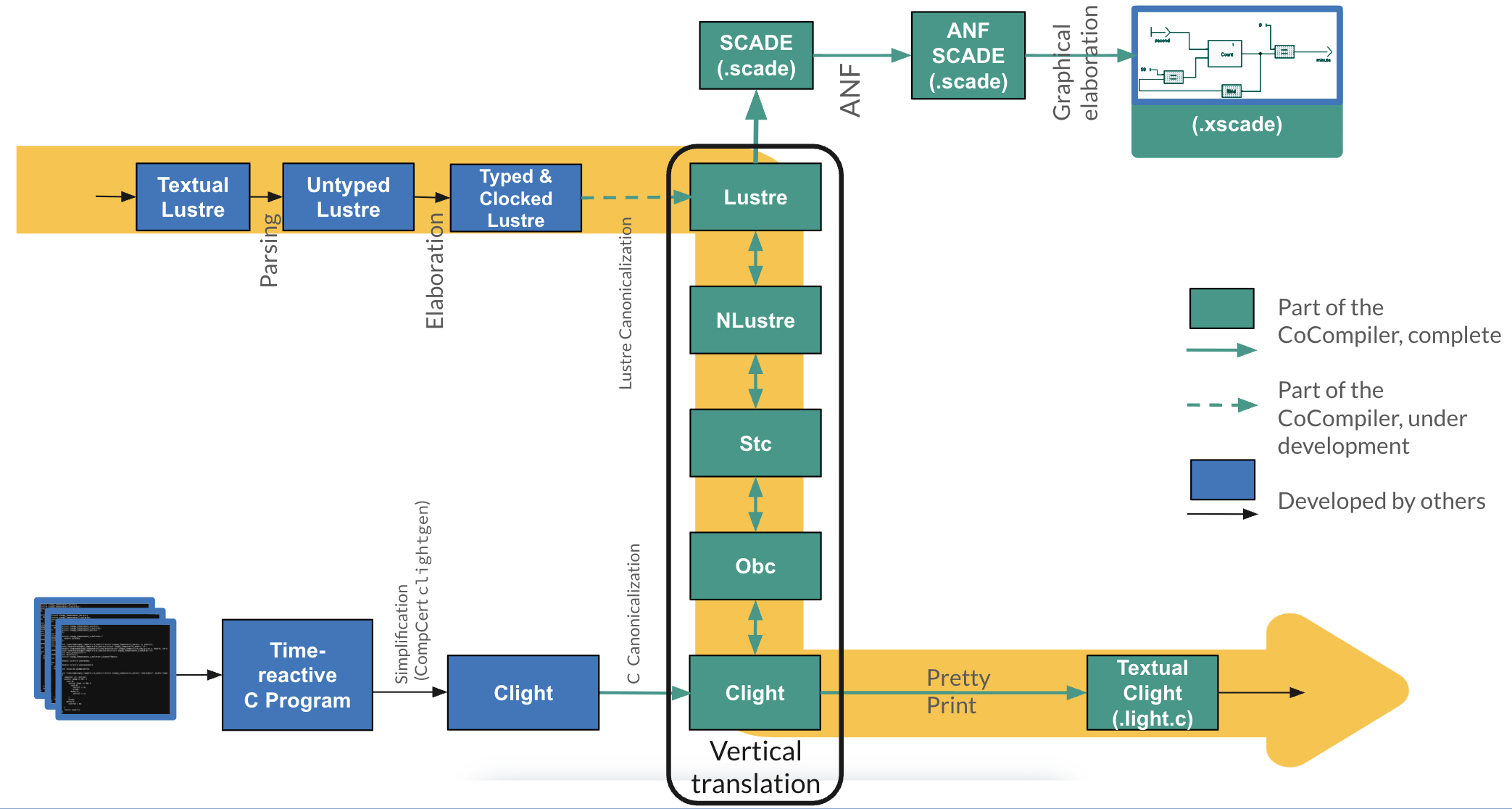}
\caption{\OurToolCap{} used as a compiler}
\label{fig:cocoCompiler}
\end{figure}

\begin{figure}
\centering
\includegraphics[scale=0.42]{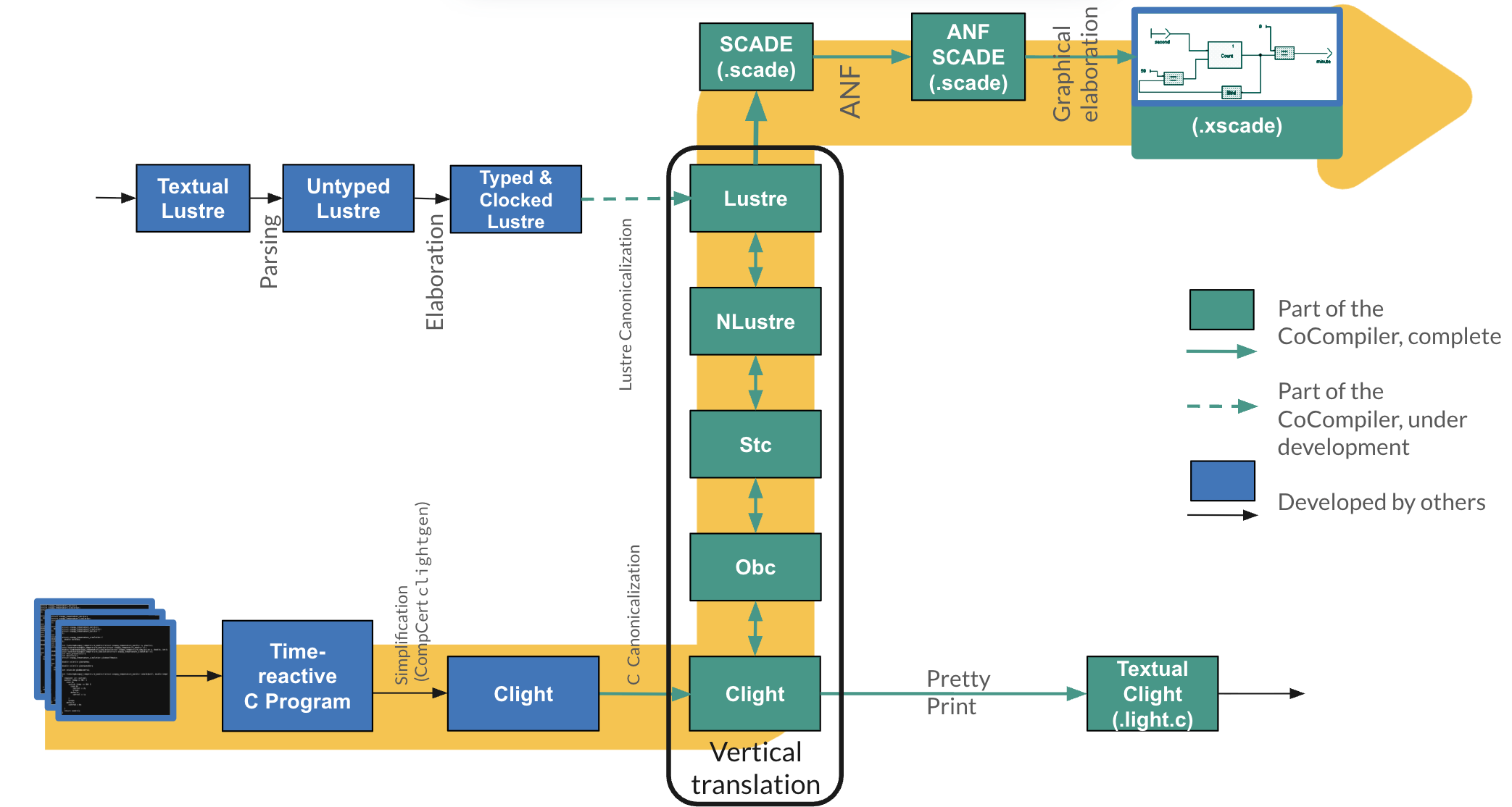}
\caption{\OurToolCap{} used as a lifter}
\label{fig:cocoLifter}
\end{figure}

The full toolchain for use of
\OurTool{}, depicted in Figures~\ref{fig:cocoCompiler}~and~\ref{fig:cocoLifter}, consists of three things:
\begin{itemize}
\item A relational bidirectional compiler from Lustre~$\leftrightarrow$~Clight
\item A Lustre~$\rightarrow$~C pipeline that includes parsing, elaboration, canonicalization, and a pretty printer for Clight output
\item A C~$\rightarrow$~Lustre pipeline, which includes parsing, canonicalization, and final transformation that produces SCADE graphical models
\end{itemize}

\OurToolCap{} can be used as a Lustre~$\rightarrow$~C compiler or a C~$\rightarrow$~Lustre~$\rightarrow$~SCADE lifter. The compilation toolchain (shown as a yellow arrow in  Figure~\ref{fig:cocoCompiler}) starts by feeding a Lustre program into an existing parser implemented in Haskell \cite{lustreParser}. We are still completing the bridge between the parser output and our Lustre AST, which will also serve as a canonicalization pass.

The lifting toolchain, shown in Figure~\ref{fig:cocoLifter}, starts with CompCert's \clightgen{} tool, which parses C programs and translates them to Clight \cite{clightGenRepo}. We then apply a set of semantics-preserving horizontal transformations to canonize the program, after which it can be lifted through the vertical relation into Lustre. These transformations are simple, mechanical rewrites—for example, re-associating sequences of statements or inserting no-op \texttt{skip} commands—and are implemented in non-relational Haskell. Despite these efforts, \OurTool{} remains sensitive to superficial variations in C syntax. Minor changes such as alpha-renaming variables or reordering function declarations can prevent successful lifting. We discuss limitations and possible remedies in Section~\ref{sec:conclusion}.

We added a single-direction translation from Lustre to graphical SCADE. This allows \OurTool{} users to lift their reactive C directly into a model that can be immediately loaded into Ansys SCADE to produce a block diagram. Our SCADE translation involves three single-direction passes. First, we translate from Lustre to SCADE. Then, there is a normalization pass resulting in ANF (Administrative Normal Form) SCADE. Finally, we translate the \texttt{.scade} file into a format consumable by Ansys SCADE. Ansys SCADE takes the resulting \texttt{.xscade} and produces a block diagram. Like our horizontal transformations, the SCADE translation is example-driven, not comprehensive.

The core of \OurTool{} is a relational reimplementation of the Vélus compiler. Specifically, we ported four Vélus compilation phases into \OurLanguage{}: Lustre~$\leftrightarrow$NLustre, NLustre$\leftrightarrow$Stc, Stc$\leftrightarrow$Obc, and Obc$\leftrightarrow$Clight\cite{velusRepo, velusWeb}. Vélus makes extensive use of set difference, a construct that proved difficult to express relationally without native support for inequality. This challenge motivated our design and implementation of efficient disequality in \OurLanguage{}, based on lazy evaluation \cite{walrus}.

\subsection*{Lifting example}
\label{subsec:liftingExample}
Now, we show how \OurTool{} lifts some C code representing the \texttt{count} program discussed in Section \ref{sec:background}. Figure \ref{fig:count.c}  shows \texttt{count.c}, a liftable C file representing the \texttt{count} example from Section \ref{subsec:lustre}.
\texttt{count.c} looks contrived due to the limited horizontal transformation capability we have currently implemented. It is structured in a similar way to C code that results from compilation with Vélus. In particular, the odd naming scheme is taken directly from Vélus's automatic naming scheme \cite{velusRepo}.

We omit a detailed description of how to write liftable C, as it does not bear directly on \OurTool{}'s proof of our approach for DSL lifting. To understand the gist of \texttt{count.c}, it is enough to know that each liftable C program consists of three components: a state (\texttt{struct count}), a state initialization function (\texttt{fun\$reset\$count}), and a step function (\texttt{fun\$step\$count}). Together, these three components implement one state machine that iterates over time. This is Vélus's C semantics for the result of a Lustre program: a stream indexed by time \cite{velusRepo}. The \texttt{reset} function initializes the stream at timestep $0$ and the \texttt{step} function advances the stream by one timestep. \OurToolCap{} does not expect a \texttt{main} function with a loop that iterates the \texttt{step} function; the three state machine components are enough to generate corresponding Lustre.

\begin{figure}
\centering
\begin{verbatim}
struct count {
    int norm1$1;
  };

  int fun$step$count(struct count *obc2c$self, int i) {
    register int o;
    o = (*obc2c$self).norm1$1 + i; 
    (*obc2c$self).norm1$1 = o;
    return o;
  }

  void fun$reset$count(struct count *obc2c$self) {
    (*obc2c$self).norm1$1 = 0; 
    return;
  }
\end{verbatim}
\caption{\texttt{count.c}}
\label{fig:count.c}
\end{figure}

\OurToolCap{} can seamlessly lift this code into \texttt{countLifted.lus} (shown in Figure \ref{fig:countLifted.lus}). \;\texttt{countLifted.lus} is written in a Vélus subdialect of Lustre called normalized Lustre; it's not as readable as the hand-written Lustre in Figure \ref{fig:count.lus}, but it is semantically equivalent \cite{velusN}. Finally, Figure \ref{fig:count-SCADE} shows the SCADE block diagram \OurTool{} generated by lifting \texttt{count.c}.

\begin{figure}
\centering
\begin{verbatim}
node count (i : int32) returns (o : int32)
  var norm1$1 : int32;
  let
    o = norm1$1 + i;
    norm1$1 = 0 fby o;
  tel
\end{verbatim}
\caption{\texttt{countLifted.lus}}
\label{fig:countLifted.lus}
\end{figure}

\section{Conclusion and future work}
\label{sec:conclusion}
In this short paper, we presented \OurTool{}, a bidirectional compiler and lifter between C and Lustre, a functional language for reactive systems. We built the bulk of \OurTool{}'s lifting and compiling functionalities by porting an existing functional DSL compiler, Vélus, to the relational setting. This is the uniqueness of \OurTool{}'s approach among DSL lifters: by writing a DSL compiler as a relation, we got a DSL lifter “for free”. This simple idea was fast to implement and is sufficient to successfully lift a canonical sublanguage of C into Lustre. In order to lift more real-world C, \OurTool{} also  applies semantics-preserving single-direction canonicalization passes. After lifting to Lustre, a final optional single-direction pass produces a graphical SCADE model of the lifted code. 

Today, \OurTool{} is only a proof of concept of our relational approach to quick and easy DSL lifting. We’ve demonstrated that repurposing compilers is a practical, efficient choice for building DSL lifters; a natural next step is to apply this approach to other DSLs. We think that lifting from a relatively high-level DSL into a higher one is an even better use-case for our relational approach, as we will not need to capture low-level compiler optimizations or intricate arithmetic reasoning in the relational setting. We are particularly interested in SysML, a popular language for modeling and specifying systems \cite{sysML}. SysML is extremely abstract and produces visual models that cannot execute; systems engineers often prefer to specify systems only at the most abstract level in SysML, rather than writing the details of executable Lustre code. It would therefore help systems engineers reason about Lustre code if they could lift it into a SysML model. On the other hand, it would also be useful for engineers to write SysML models and compile them down to Lustre to get even a partially completed model that can execute. Because Lustre and SysML are both relatively high level and because both the compilation and lifting directions are of interest to Lustre and SysML users, we would like to extend \OurTool{} with a C~$\leftrightarrow$~Lustre~$\leftrightarrow$~SysML pipeline. This new feature would complement our existing C~$\rightarrow$~Lustre~$\rightarrow$~SCADE lifting capability.

\OurToolCap{}'s existing lifting functionality could also be improved. The vertical translation would benefit from the addition of more Lustre features. \OurToolCap{}'s development was example driven; as a result, some more advanced Lustre features, such as \texttt{merge} and \texttt{reset}, are not currently supported. Additionally, we hope to add Vélus’s advanced features, such as normalization, scheduling, or optimization, to the vertical translation. As discussed in Section \ref{subsec:implementation}, \OurTool{} is brittle when encountering minor structural differences in C files. Adding more horizontal passes to canonicalize a broader swath of C code would make \OurTool{} a more practical DSL lifter. We would also like to make \OurTool{} more useful to systems engineers by improving its ability to translate Lustre into SCADE block diagrams, either by expanding our own single-direction translation or by connecting \OurTool{} to an existing Lustre~$\rightarrow$~SCADE lifter. 
Finally, \OurTool{} users have expressed that the automatically generated SCADE diagrams, while correct, are harder to read than hand-made diagrams. Up until now, we have targeted correctness of lifted code and breadth of liftable code. We have ideas for how to improve readability of lifted code and block diagrams, including Lustre de-normalization, changing our automatic variable naming scheme, and specifying the layout of the generated SCADE blocks.

\bibliography{sources.bib} 

\end{document}